# A plug-and-play molecular approach for room temperature polariton condensation

Prathmesh Deshmukh[1,2], Sitakanta Satapathy[1*], Evripidis Michail[3], Andrew H. Olsson[4], Rezlind Bushati[1,2], Ravindra Kumar Yadav[1], Mandeep Khatoniar[1,2], Junsheng Chen,[5] George John[6], Bo W. Laursen[5], Amar H. Flood[4], Matthew Y. Sfeir[2,3] & Vinod M. Menon[1,2*]

*[1]Department of Physics, Center for Discovery and Innovation, The City College of New York, 85 St. Nicholas Terrace, New York, NY 10031, USA.*

*[2]The PhD Program in Physics, The Graduate Center of the City University of New York, 365 5th Ave, New York, NY, 10016, USA.*

*[3]Photonics Initiative, Advanced Science Research Center, City University of New York, New York, New York 10031, USA*

*[4]Department of Chemistry, Indiana University, Bloomington, IN 47405, USA*

*[5]Nano-Science Center and Department of Chemistry, University of Copenhagen, Denmark*

*[6]Department of Chemistry, Center for Discovery and Innovation, The City College of New York, 85 St. Nicholas Terrace, New York, NY 10031, USA.*

Exciton-polaritons (EP), half-light half-matter quasiparticles that form in optical cavities, are attractive platforms for creating macroscopic coherent states like BECs. EPs based on organic molecules are of particular interest for realizing such states at room temperature while offering the promise of synthetic tunability. However, the demonstrations of such condensates have been limited to a few specific molecular systems[1]. Here we report a universal platform for realizing molecular polariton condensates using commercial dyes that solves long standing material challenges. This solution is made possible using a new and programable molecular material called small-molecule, ionic isolation lattices (SMILES) with the potential to incorporate a wide array of molecular fluorophores[2]. We show EP condensation in rhodamine by incorporating it into SMILES lattice placed in a planar microcavity. The SMILES approach overcomes the major drawbacks of organic molecular photophysical systems such as self-quenching, which sets the foundation for realizing practical polaritonic devices operating at ambient temperatures covering wide spectral range**.**

* Corresponding Authors: ssatapathy@ccny.cuny.edu ; vmenon@ccny.cuny.edu

EP condensates have become a platform for both fundamental investigations and technological developments[3,4]. Owing to their smaller effective mass, EPs have the advantage of realizing Bose Einstein condensation at elevated temperatures. While not true BECs since they are not at thermal equilibrium, they exhibit some of the novel features like superfluidity[7,8] and, by analogy to cold atom lattices, can be implemented as Hamiltonian emulators[9]. These properties motivate exploration and critical evaluation of new optical materials to support EP condensation. To this end, we report the first use of small-molecule, ionic isolation lattices (SMILES, Fig 1a)[2] that promise to incorporate commercially available organic fluorophores directly and easily into optical devices for EP condensation.

There have been different materials platforms that show polariton condensation at room temperature such as wide band gap semiconductors, 2D materials, perovskites and organic molecular solids[1,10–12]. While early experiments used inorganic semiconductors hosting Wannier excitons[5], for over a decade now there has been much interest in realizing condensation using large binding energy Frenkel excitons in organic molecular solids[6]. The organic molecular systems have garnered much attention owing to the promise of on-demand synthetic tunability, whether for materials compatibility or control of photophysical properties, as well as their potential integration with flexible substrates. Despite this promise, the demonstrations to date have been carried out using only a few specific systems ranging from single crystals to ladder type polymers and fluorescent proteins[1,3–11]. It would be ideal if we could instead choose from the large body of organic fluorescent dyes available today to truly unleash the synthetic power of organic fluorophores.

Realization of EP condensation from any arbitrary dye has not previously been conceivable owing to the unavoidable and deleterious effects of a few different but coupled processes. First, exciton-exciton annihilation inhibits access to the densities required to surpass the condensation condition to generate macroscopic coherence where the inter particle separation is less than their de Broglie wavelength. Second, concentration quenching makes it difficult to realize thin films of dyes at the molecular densities needed in solid matrices for polariton generation without compromising drastically on the photo luminescence quantum yield. Third, the inherent structural disorder emerging when dyes aggregate at higher concentrations impedes large area condensates needed for applications such as polariton-based computations and signal processing. A common solution to all these issues is to dilute the dyes and hence using devices composed of thick organic films.

The recent discovery of fluorescent SMILES material[12] has the potential to address these challenges in a straightforward way. Concentration quenching arising from dye aggregation is a long-standing problem that impacts the vast majority of the 100,000+ organic dyes in existence and constitutes a general bottleneck in materials creation for many solid-state photonic applications. SMILES provide a universal solution to the problem of aggregation by isolating the dyes ~15 Å from each other within high-density solids. SMILES are readily formed (Fig 1a) by mixing cationic dyes with the colorless and anion-binding macrocycle, called cyanostar (CS) [2]. The macrocycles capture the counter anions associated with the dyes to produce negatively charged building blocks that stack alternately with the dyes in charge-by-charge lattices[13] . Major classes of commercial dyes, including xanthenes, oxazines, styryls, cyanines, and trianguleniums, show fluorescence turn-on in solid state SMILES materials resulting in the brightest known emitters (volume-normalized brightness > 7,000 $M^{-1}$ $cm^{-1}/nm^3$) as solids, crystals, thin films, nanoparticles and doped polymers[12,14–16]. Simple preparation, high photo luminescence quantum yields, and dye-dye isolation promise to circumvent all the issues of annihilation, quenching and aggregation when organic dyes are dissolved into polymer films.

Here we demonstrate the formation of room temperature polariton condensation using rhodamine 3B (R3B), a widely used laser dye (with bandgap ~ 2.19 eV, ~ 566 nm), composed as a SMILES material. We used the R3B dye as the perchlorate ($ClO_4^-$) salt, combined it with two molar equivalents of cyanostar and dissolved it in PMMA at a final dye concentration of 10 mM. A thin film of the polymer-SMILES material was produced by spin coating for integration into a Fabry Perot microcavity structure to demonstrate condensation. The R3B-SMILES film was compared to PMMA-based film incorporating the R3B perchlorate alone (10 mM). Our prior work suggests that the SMILES materials will form charge-by-charge assemblies within the PMMA. The resulting isolation of the R3B dyes in the SMILES based films are highly monomeric.  A ~10× enhancement in photoluminescence (PL) quantum yield is observed compared to the dye alone and suggests the possibility to implement thinner films than previously studied in the context of molecular polariton condensates[1,7].

The brightness and photostability of thin PMMA films (Fig. 1b) composed of R3B-SMILES is superior to those of the R3B alone. The tolerance to photo-damage under laser illumination and the reduction in exciton-exciton annihilation of the SMILES films are tested using a 10 kHz pulsed laser at 530 nm wavelength. The saturation threshold of R3B-SMILES is found to be five times higher than R3B by itself indicating reduced quenching and photo-degradation.

Fortuitously, we observed favorable orientational ordering of the molecular transition dipole moments in the thin PMMA films of R3B-SMILES. Fourier imaging microscopy (FIM) was used to image the emission dipole using PL measurements[17]. The observed Fourier space PL maps observed at the back focal plane (Fig. 1c, 1d) along with the line cut along the $k_x$ direction (Fig 1e) shows the in-plane orientation of the dipoles in the SMILES film. The emission pattern of the R3B SMILES (blue) with maximum emission at $k_y = 0$ is indicative of the dipoles being aligned in-plane in contrast to the R3B dye (magenta) showing emission pattern indicative of randomly oriented dipoles[17]. The in-plane orientation of the transition dipole moments in the R3B SMILES film enhances the coupling of the excitons to the cavity photons in a planar Fabry-Perot cavity. Thus, in addition to the photostability and high quantum yield, the ideal dipole orientation provided by the SMILES material offers enhanced light-matter interactions.

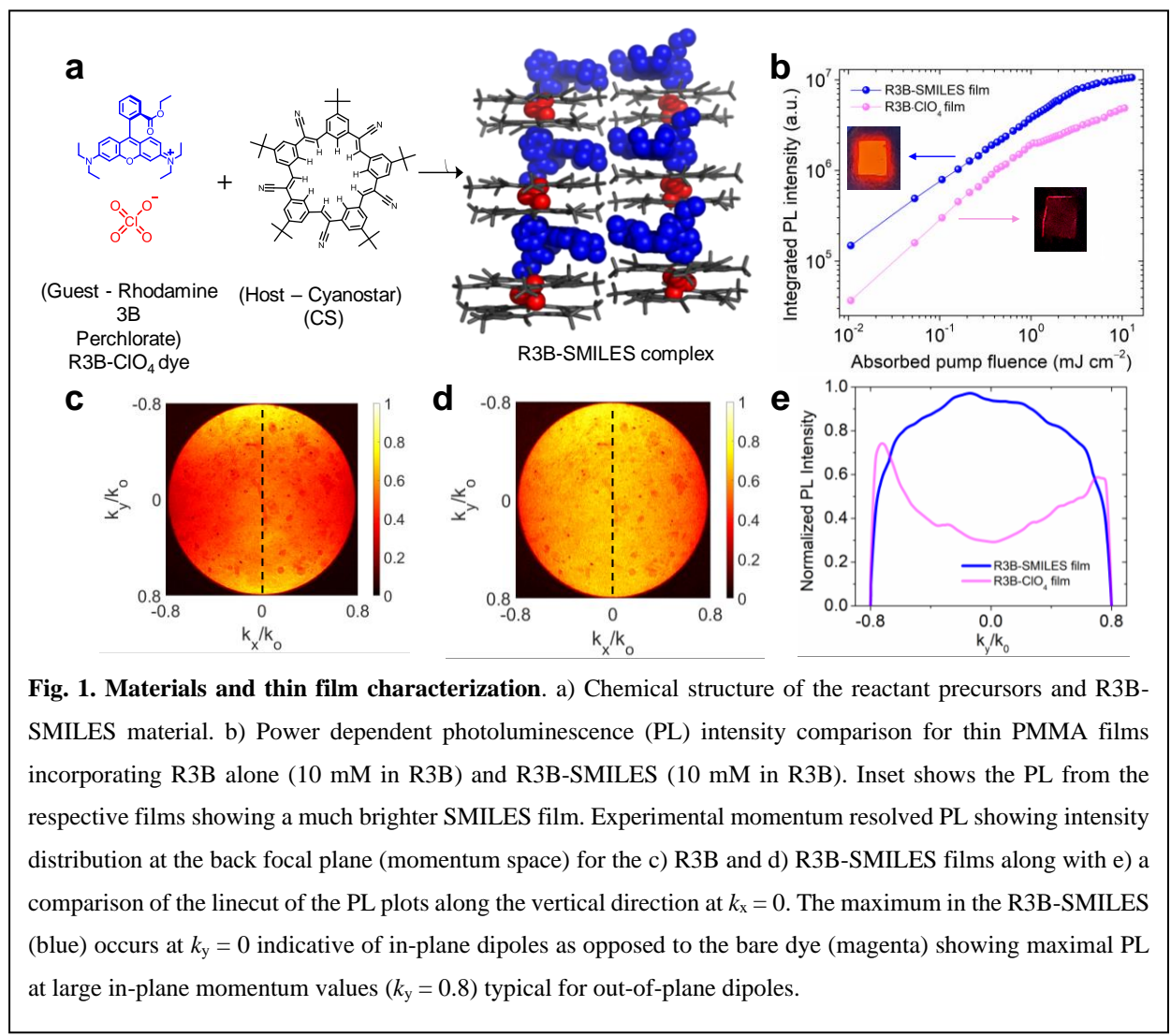

**Fig. 1. Materials and thin film characterization**. a) Chemical structure of the reactant precursors and R3B-SMILES material. b) Power dependent photoluminescence (PL) intensity comparison for thin PMMA films incorporating R3B alone (10 mM in R3B) and R3B-SMILES (10 mM in R3B). Inset shows the PL from the respective films showing a much brighter SMILES film. Experimental momentum resolved PL showing intensity distribution at the back focal plane (momentum space) for the c) R3B and d) R3B-SMILES films along with e) a comparison of the linecut of the PL plots along the vertical direction at $k_x = 0$. The maximum in the R3B-SMILES (blue) occurs at $k_y = 0$ indicative of in-plane dipoles as opposed to the bare dye (magenta) showing maximal PL at large in-plane momentum values ($k_y = 0.8$) typical for out-of-plane dipoles.

The optical cavity (Fig. 2a) is composed of a spin coated ~ 35 nm thin PMMA film of R3B-SMILES (see Methods) sandwiched between a silver top mirror and a bottom $SiO_2/TiO_2$ distributed Bragg reflector (DBR) on a quartz substrate. The resulting structure supports a Tamm plasmon mode[18,19] with field maximum overlapping with the spatial position of the SMILES layer as shown in Supplementary Fig. S2. The Tamm plasmon cavity showed a loaded $Q$-factor of ~ 150. The experimental white-light reflectivity obtained from the cavity sample using FIM is shown in Fig. 2b. Strong coupling of a Tamm cavity mode (2.13 eV) with the main molecular exciton of R3B-SMILES ($S_{01}$ at 2.19 eV, as shown in the absorption spectrum plotted as a grey line in the left panel of Fig. 2c) resulted in two polariton branches with Rabi splitting, $\hbar\Omega_R \sim 100$ meV. A similar microcavity comprised of the same concentration and thickness of the R3B dye alone in the PMMA film as active layer resulted in weak coupling (Fig. S1) despite the negligible difference in excitation and emission energies compared to R3B-SMILES. This difference in performance is attributed to the synergy between high absorption strength and predominantly in-plane dipole orientation in the R3B-SMILES observed from the thin PMMA film.

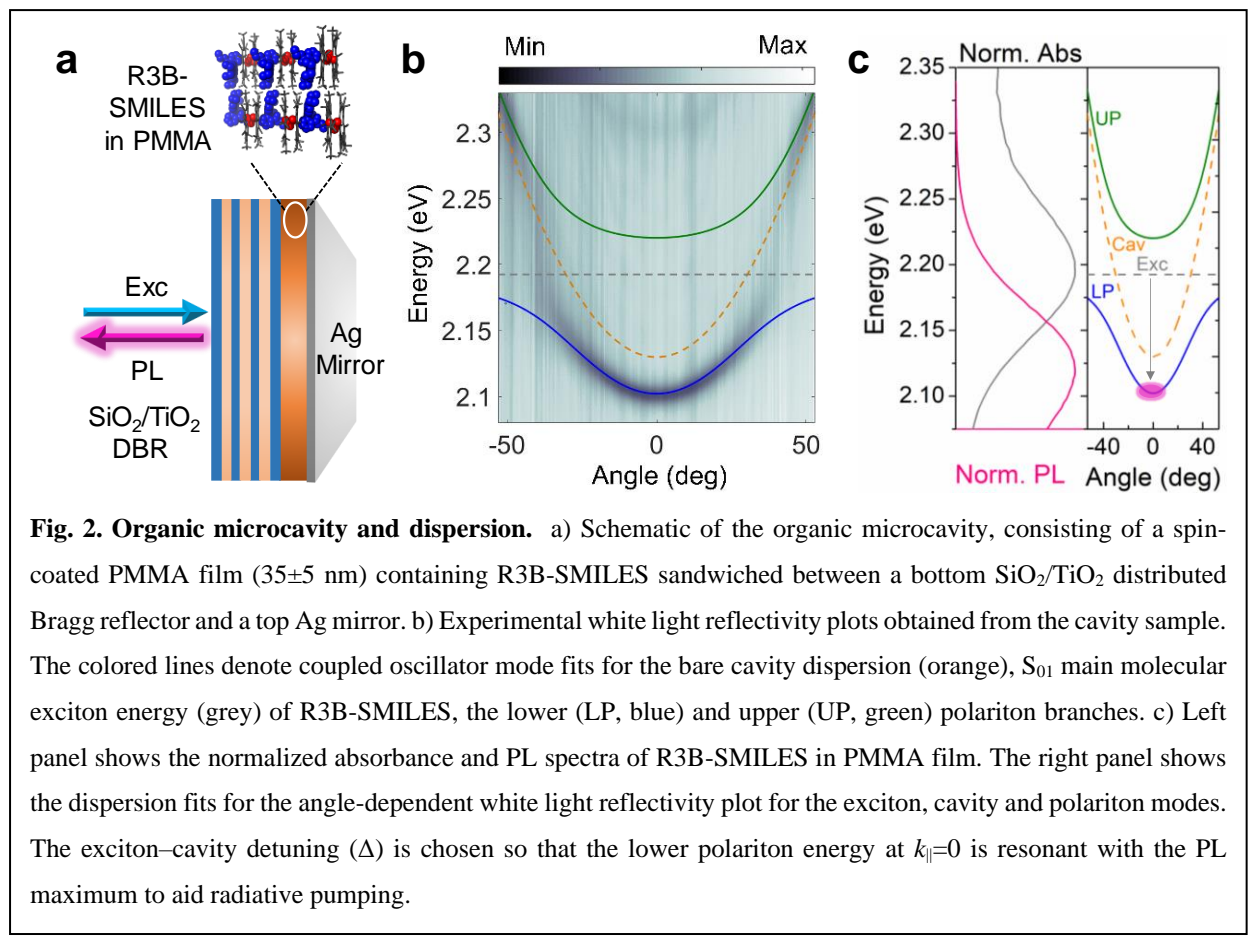

**Fig. 2. Organic microcavity and dispersion.** a) Schematic of the organic microcavity, consisting of a spin-coated PMMA film (35±5 nm) containing R3B-SMILES sandwiched between a bottom $SiO_2/TiO_2$ distributed Bragg reflector and a top Ag mirror. b) Experimental white light reflectivity plots obtained from the cavity sample. The colored lines denote coupled oscillator mode fits for the bare cavity dispersion (orange), $S_{01}$ main molecular exciton energy (grey) of R3B-SMILES, the lower (LP, blue) and upper (UP, green) polariton branches. c) Left panel shows the normalized absorbance and PL spectra of R3B-SMILES in PMMA film. The right panel shows the dispersion fits for the angle-dependent white light reflectivity plot for the exciton, cavity and polariton modes. The exciton–cavity detuning (Δ) is chosen so that the lower polariton energy at $k_{\parallel} = 0$ is resonant with the PL maximum to aid radiative pumping.

An important consideration in our design of the Tamm cavity was the detuning between the Tamm plasmon mode and the excitonic transition, $\Delta = \omega_{cav} - \omega_{ex}$ at $k_\parallel = 0$. The detuning was designed so that the lower polariton at $k_\parallel = 0$ was resonant with the PL emission maximum of the dye molecule thereby ensuring radiative pumping of the lower polariton from the exciton reservoir, a technique that has been shown to be highly efficient in realizing polariton condensation in organic molecular systems[6,20,21]. The detuning was controlled by varying the thickness of the SMILES films.

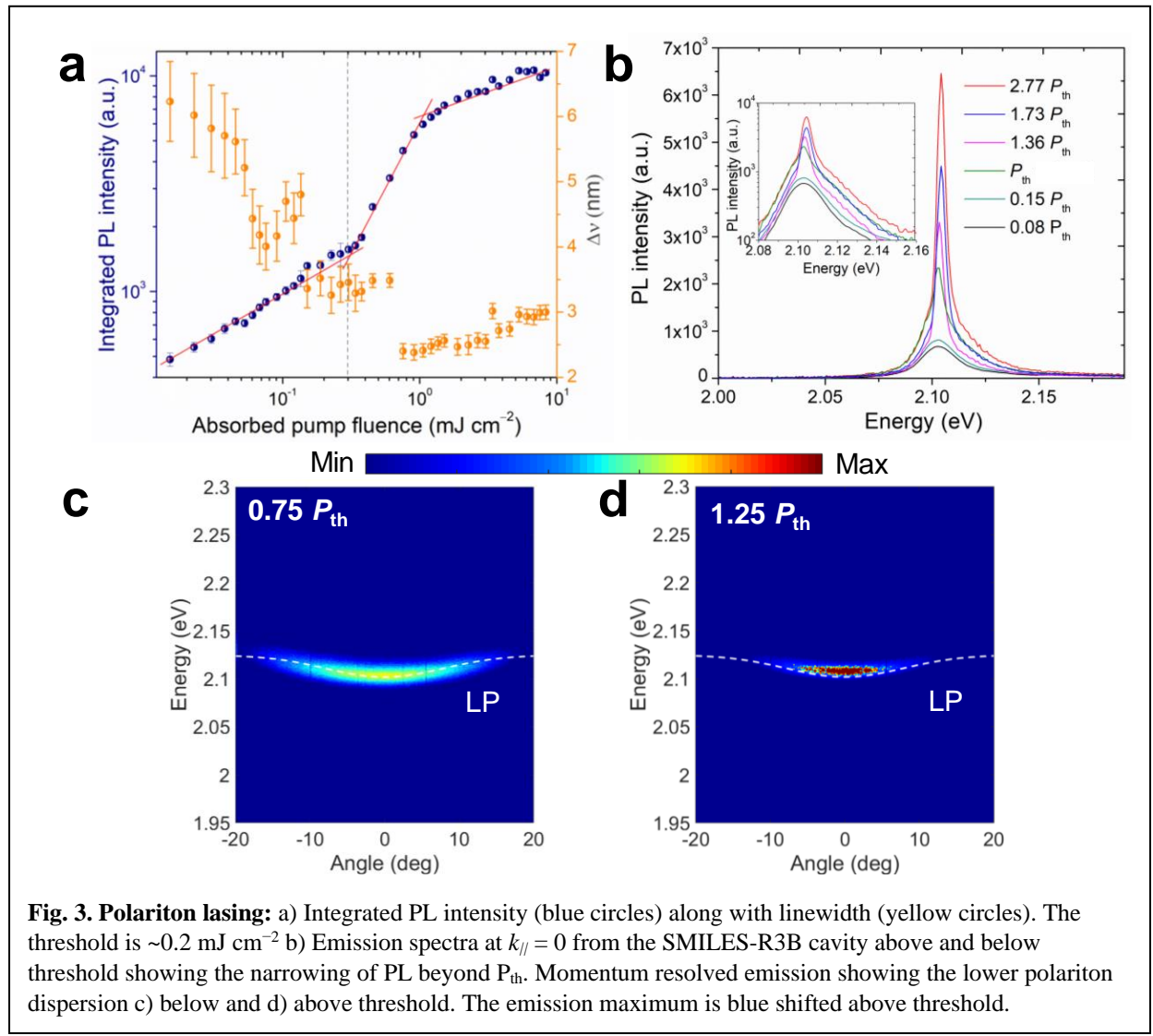

**Fig. 3. Polariton lasing:** a) Integrated PL intensity (blue circles) along with linewidth (yellow circles). The threshold is ~0.2 mJ cm$^{-2}$ b) Emission spectra at $k_{//} = 0$ from the SMILES-R3B cavity above and below threshold showing the narrowing of PL beyond $P_{th}$. Momentum resolved emission showing the lower polariton dispersion c) below and d) above threshold. The emission maximum is blue shifted above threshold.

Following the demonstration of strong coupling and production of EPs in the R3B-SMILES cavity systems, we carried out studies to determine the onset of condensation. For this purpose, we quantify how luminescence responds to pump fluence. The cavity was pumped at 2.34 eV, corresponding to the blue edge (reflectivity minimum at normal incidence) of the DBR using a 150-fs laser operating at 10 kHz (see Methods for details). The emission was imaged in the momentum space (Fig. 3c and 3d) to study how the dispersion is modified under high polariton densities. The integrated PL intensity from the LP branch was studied as a function of pump fluence (Fig 3a, blue dots, red lines). Above the threshold for EP condensation at 0.2 mJ cm$^{-2}$ (grey vertical line) a super linear increase in PL intensity is observed along with a collapse in the linewidth ($\Delta\nu$, black dots) indicative of the increase in temporal coherence of the emission. Beyond this threshold, we find a slow power-dependent linewidth broadening. This behavior has been discussed extensively in previous reports and has been attributed to phase diffusion stemming from polariton self-interaction within the condensate as well as to condensation within different localized modes[5,22–24]. The individual spectra, measured at normal incidence at $k_{//} = 0$, are shown in Fig. 3b. Emission from the LP states show distinct narrowing and almost an order of magnitude increase in peak intensity above condensation threshold[17]. The dispersion of the polariton emission mapped using FIM, below and above the threshold is shown in Fig. 3c and 3d, respectively. Above the threshold we observe a collapse in momentum space along with a blue shift (Supplementary Fig. S3) providing another indication of onset of polariton condensation. The observed threshold is ~ 0.2 mJ cm$^{-2}$, which is comparable to those reported for dye-based cavities with higher $Q$-factors[7]. The observed blue shift in energy of the PL peak above the condensation threshold (seen in the k-space image Fig. 3d and Fig. S3) results from the combination of two effects[9,25]: (i) saturation-induced quenching of the exciton-photon coupling, thus lowering the Rabi splitting energy and (ii) renormalization of the cavity mode energy stemming from the change in effective refractive index ($n_{eff}$) of the cavity when exciting weakly coupled or uncoupled molecules. We also observed the real space emission spot narrowing above threshold (Supplementary Fig. S4). This narrowing is attributed to the local disorder in organic films that results in localized emission from LP as has been reported previously in such systems[5].

The polariton condensation observed in the R3B-SMILES microcavity was achieved under ambient conditions as expected from the large binding energy of Frenkel excitons. The maximum repetition rate of the 2.34 eV pump at which condensation was observed of 10 kHz, is one of the highest reported among organic systems. As seen with the pump-dependent study (Fig 1c), the tolerance to higher repetition rate reinforces the advantage of the SMILES platform against the range of processes that typically deteriorate condensation (pump laser induced heat, $O_2$ induced bleaching, bimolecular annihilation).

The polarization dependence of the LP emission below (0.75 $P_{th}$) and above (1.25 $P_{th}$) threshold are shown in Fig. 4a. The polarization of the emission is observed to be pinned to the polarization of the excitation laser only above the condensation threshold (magenta) with a degree of linear polarization of ~ 73%. Shown in Fig. S5 is the emission spectrum collected at different polarizer angles showing clearly the linearly polarized nature of emission from the EP condensate. Similar reports of linearly polarized emission from organic polariton condensates in the past have been attributed to preferential absorption and emission of a specific polarization[5].

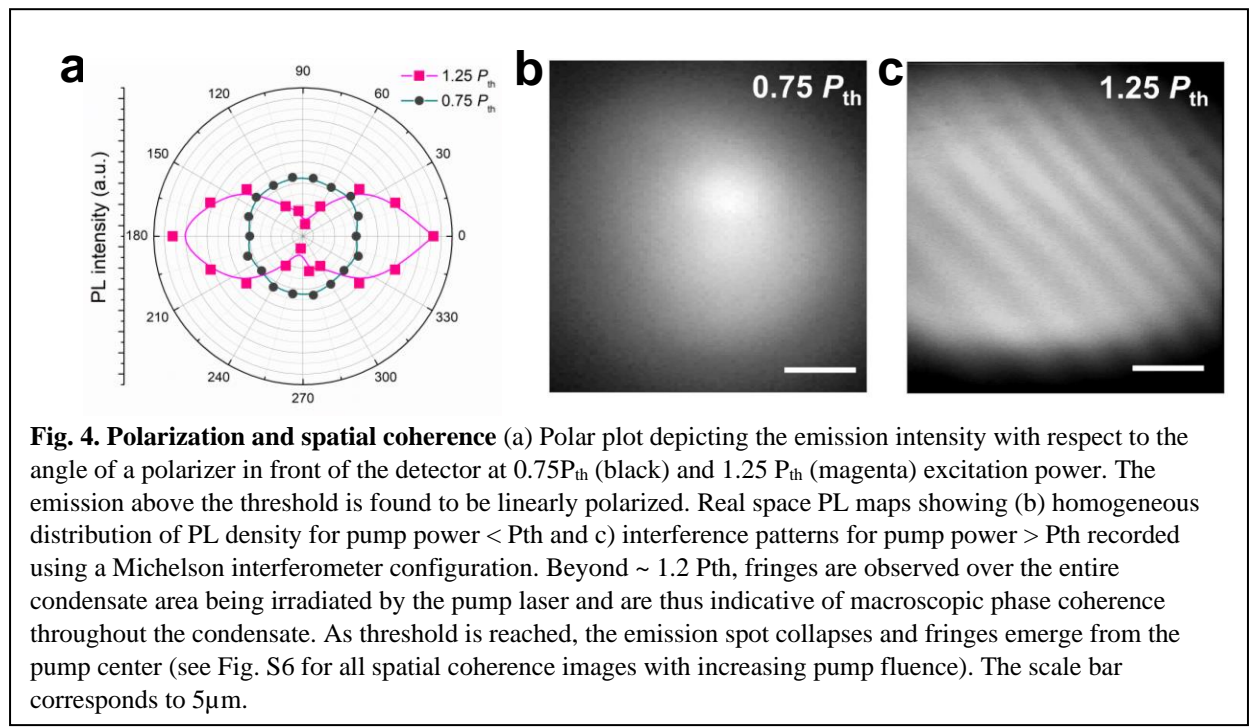

**Fig. 4. Polarization and spatial coherence** (a) Polar plot depicting the emission intensity with respect to the angle of a polarizer in front of the detector at 0.75$P_{th}$ (black) and 1.25 $P_{th}$ (magenta) excitation power. The emission above the threshold is found to be linearly polarized. Real space PL maps showing (b) homogeneous distribution of PL density for pump power < Pth and c) interference patterns for pump power > Pth recorded using a Michelson interferometer configuration. Beyond ~ 1.2 Pth, fringes are observed over the entire condensate area being irradiated by the pump laser and are thus indicative of macroscopic phase coherence throughout the condensate. As threshold is reached, the emission spot collapses and fringes emerge from the pump center (see Fig. S6 for all spatial coherence images with increasing pump fluence). The scale bar corresponds to 5µm.

One of the defining features of condensate formation is the spontaneous appearance of long-range order in the form of spatial coherence.[23,26] This is usually established using an interferometric technique[23]. Here we use a Michelson interferometer in the retroreflector geometry to study the spatial coherence below and above the threshold[27]. Shown in Fig. 4b (below threshold) and 4c (above threshold) are the interference profiles showing clearly the interference pattern above threshold indicating the onset of spatial coherence in the polariton emission. Below threshold (0.75 $P_{th}$), the polariton's PL intensity appears homogeneously distributed over the entire pump spot. Interference fringes at $P_{th}$ begin to emerge (Fig. S6) consistent with the onset of coherent emission from cavity polaritons. With increase in the pump fluence to > 1.2 $P_{th}$, fringes are readily identified over the entire condensate area illuminated by the pump laser (spot size = 12µm), which clearly demonstrates the presence of long-range order. [28–32]

In summary, we demonstrated a platform for realizing polariton condensates at room temperature using commercially available organic dyes based on SMILES materials. As an indication of the potential for a universal plug-and-play platform, we used rhodamine 3B perchlorate asan archetypical organic laser dye. The thin PMMA films of R3B-SMILES were found to be highly photostable showing strong tolerance to pump fluence along with in-plane dipole orientation facilitating better coupling to the electromagnetic field in the cavity. The optically-active PMMA film of R3B-SMILES was integrated by spin-coating into a metal-DBR hybrid microcavity with a moderate Q-factor (~150) leading to the formation of polariton condensates. The large spatial coherence and spatial uniformity in the emission points toward low defect density in the R3B-SMILES films adding to the advantages of the SMILES platform for polariton condensation.

Given that SMILES platform has been shown to work with a wide variety of commercial organic dyes to fully express their bright luminescence in solid state[12], we foresee the extension of the present work to the large class of commercially available organic cationic dyes. Our results hold the promise for achieving a universal solution to the long-standing problem of an optimal emitter and microcavity design to achieve on-demand and room temperature exciton-polariton condensates spanning a wide spectral range. We anticipate these results will make organic polariton condensates more viable for realizing polaritonic devices such as Hamiltonian simulators, femtojoule switches and for modifying chemistry in the condensate regime.

**Methods**

*Formation of R3B-SMILES in host PMMA:* The synthesis of cyanostar has been reported[2]. To prepare the films, a PMMA C2 solution (2% w/w PMMA in chlorobenzene, Kayaku Chem) was diluted by a factor of two using chlororobenzene. To this solution was added cyanostar and R3B•ClO₄ (Exciton, Luxottica. Inc.) to make the final concentration of 20 mM and 10 mM, respectively. The solution is subjected to vigorous stirring for 48 hours at room temperature to undergo formation of the R3B-SMILES assemblies within the PMMA solution. Preparation of solutions of the dye alone involved addition of 10 mM of R3B•ClO₄ to 1 mL of the PMMA C2 solution followed by vigorous stirring for 48 hours at room temperature.

*Fabrication of thin films and microcavity samples:* The SiO₂/TiO₂ DBR substrates were subjected to O₂ plasma treatment for ~ 5 mins, which helped to produce better spreading of the R3B-SMILES PMMA solution prior to spinning. The R3B-SMILES solution was then spin-coated using a three-step recipe. First, about 70 µL of the solution was delivered to the O₂ plasma treated DBR substrate. Second, the substrate was spun at 2000 RPM (*t* = 15 sec) and then 4000 RPM (*t* = 120 sec). These films were subjected to drying in vacuum under constant pressure (2500 Pascal) in the dark at room temperature for ~48 hrs. This processing step was found to generate a uniform film thickness across a large area (~1500-2000 µm² along the horizontal plane) estimated from the direct method using a Profilometer (Bruker Dektak-XT). The thickness value estimated from profilometry was ~ 35 nm. All these films are then coated with the top mirror by slow evaporation of silver (0.2 Å / sec) using the e-beam evaporation technique. We deposited 100 nm of silver mirror on the top for completing the cavity.

*Linear Optical Spectroscopy*: Measurements of the angle-resolved, white light reflectivity were conducted using emission from a broad band tungsten halogen lamp. A telescope (1:1 arrangement) with a 100 µm pinhole at the beam waist was used to collimate the beam. The reflected signal was collected using a 50×, 0.8 numerical aperture (NA) objective (same as the incident beam) followed by deflection from a beam splitter to the spectrometer. Solid-state UV-visible measurements on quartz slides were carried out using a Jasco-760 UV-visible spectrophotometer. Angle-resolved PL measurements were performed using a homemade setup comprising a laser coupled with a Princeton Instruments monochromator with a PIXIS: 256 electron-multiplying charge-coupled device (EMCCD) camera. For both momentum-resolved white light and PL measurements, the back focal plane of the imaging objective was projected onto the EMCCD for single shot dispersion collection. For transient dipole imaging (Fig 1c, 1d); the entire Fourier circle from the imaging lens was projected onto the EMCCD in an open slit configuration to map in-plane momenta $k_x$ and $k_y$. A narrow spectral linewidth was chosen for this imaging around the center emission maxima ($\pm$ 10 nm) for better contrast.

*Pump Power-dependent Photoluminescence Measurements:* The pump beam (530 nm) was generated in a optical parametric amplifier (Light Conversion) pumped by the 1030 nm output of an amplified Yb:KGW laser (Light Conversion Carbide), and the pump beam was focused onto the sample using the 20x objective (NA = 0.40). Our *k*-space setup (with the 10 kHz repetition rate laser) designed for the power-dependent PL experiments was aligned and calibrated with this microscope objective (20x, 0.40 NA) to monitor the collapse of PL intensity around the lower (ground state) *k* vectors. Photoluminescence emission was collected in a reflection configuration using the spectrometer (Princeton Instruments, Acton SpectraPro SP-2500) and charge-coupled device (CCD) camera (Princeton Instruments, PIX 1024B). A 40% DBR transmission (averaged along all angles of incidence) and ~ 25% R3B-SMILES absorption was included in estimating the absorbed fluence values at 530 nm (2.3eV) pump. The residual excitation beam was blocked using a 550 nm long-pass filter. The real space measurements were carried out with an open slit configuration.

*Spatial Interferometry:* Spatial coherence measurements were made using a Michelson interferometer configuration. The condensate emission was collected with a 20x objective (NA = 0.40) objective and directed to a thin non-polarizing beam splitter. One output arm of the beam splitter consisted of a mirror attached to a micrometer screw gauge stage, and the other consisted of a reflector mirror used to invert the image along both the vertical and horizontal axes. After returning through the beam splitter, the two nearly collimated beams were focused onto a CCD using a $f$ = 12 mm lens, thus providing a ~47$\times$ magnification. A neutral density filter was placed in one arm to equilibrate the intensities and a 550 nm long-pass filter was used to eliminate any residual pump light.

*Coupled Oscillator Model:* The coupling of the exciton to a single cavity photonic mode is represented by the following Hamiltonian matrix:[28]

$$H = \begin{pmatrix} E_{ex} - \dfrac{i\zeta ex}{2} & g \\ g & E_{cav} - \dfrac{i\zeta cav}{2} \end{pmatrix}$$

Wherein *g* is the coupling strength of the cavity mode with the excitonic resonance and gives the Rabi splitting of interaction as stated in the main text. Both $\zeta ex$ and $\zeta cav$ are the linewidths (FWHM) of the exciton and cavity modes. We solve for the Eigen modes of the above Hamiltonian to get the polariton branch dispersion as a function of in-plane momentum (angles) and fit these with our experimental data for upper, and lower polaritonic branches. We also extract Hopfield coefficients from the same interacting Hamiltonian as shown below:

$$|UPB\rangle = C(k)|exciton\rangle + X(k)|photon\rangle$$

$$|LPB\rangle = X(k)|exciton\rangle - C(k)|photon\rangle$$

Here, *C(k)* and *X(k)* are Hopfield coefficients describing the photon and exciton fraction in the upper (UPB) and lower polariton branches (LPB).

**Supporting Information**

Supporting Information is attached to this main manuscript.

**Acknowledgments**

The authors acknowledge the use of the nanofabrication and imaging facility at ASRC, CUNY.

**Funding:** V.M.M, S.S, P.D and R.B acknowledge support from the US National Science Foundation (NSF-QTAQS program OMA-1936351) and the US Air Force Office of Scientific Research – MURI grant FA9550-22-1-0317. M.Y.S work was supported by the US Department of Energy, Office of Science, Office of Basic Energy Sciences under award no. DE-SC0022036. A.H.O and A.H.F. acknowledge support from the US National Science Foundation (DMR-2118423).

**Author Contributions:** V.M.M. and S.S. conceived the idea and designed the experiments. P.D. and S.S. contributed equally to this manuscript. S.S. made the samples including thin film optimization, cavity device fabrication, optical characterization, and Raman measurements. P.D. setup the optics for all the power dependent measurements. R.B., M.E. and M.Y.S. assisted in designing the optics. P.D., M.E. and S.S. performed the power dependent measurements with inputs from M.Y.S. S.S. and P.D. analyzed the data and wrote the manuscript with input from V.M.M. A.H.F. B.W.L. The synthesis and characterization of cyanostar and SMILES materials was performed by J.C and A.H.O. All authors discussed the results and commented on the manuscript. All authors have given approval to the final version of the manuscript.

**Conflict of Interest**: The authors declare no conflict of interest.

**Data availability:** All data will be provided by the corresponding authors upon reasonable request.

# Supporting Information

**A plug-and-play molecular approach for room temperature polariton condensation**

Prathmesh Deshmukh[1,2], Sitakanta Satapathy[1], Evripidis Michail[3], Andrew H. Olsson[4], Rezlind Bushati[1,2], Ravindra Kumar Yadav[1], Mandeep Khatoniar[1,2], Junsheng Chen,[5] George John[6], Bo W. Laursen[5], Amar H. Flood[4], Matthew Y. Sfeir[2,3] & Vinod M. Menon[1,2]

[1]*Department of Physics, Center for Discovery and Innovation, The City College of New York, 85 St. Nicholas Terrace, New York, NY 10031, USA.*

[2]*The PhD Program in Physics, The Graduate Center of the City University of New York, 365 5th Ave, New York, NY, 10016, USA.*

[3]*Photonics Initiative, Advanced Science Research Center, City University of New York, New York, New York 10031, USA*

[4]*Department of Chemistry, Indiana University, Bloomington, IN 47405, USA*

[5]*Nano-Science Center and Department of Chemistry, University of Copenhagen, Denmark*

[6]*Department of Chemistry, Center for Discovery and Innovation, The City College of New York, 85 St. Nicholas Terrace, New York, NY 10031, USA.*

[*]Corresponding Author: ssatapathy@ccny.cuny.edu; vmenon@ccny.cuny.edu

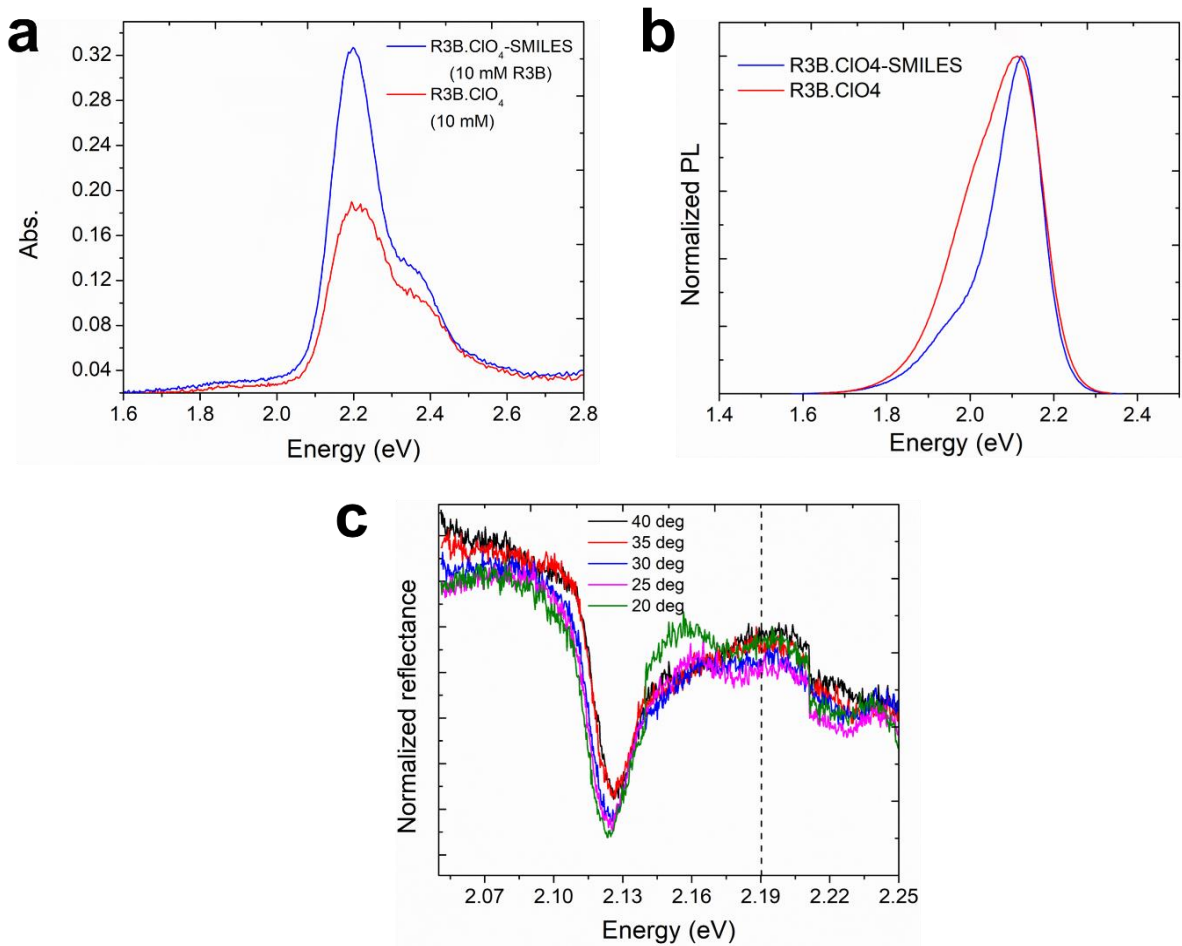

**Figure S1. a)** Absorbance of rhodamine dye (R3B.ClO$_4$) and encapsulated rhodamine dye with Cyanostar macrocycles (R3B.ClO$_4$-SMILES) for the same molecular concentration. The data shows significant enhancement of absorbance of R3B.ClO$_4$-SMILES over bare R3B.ClO$_4$ in the solid-state **b)** Self normalized photoluminescence spectra of bare R3B.ClO$_4$ and R3B.ClO$_4$-SMILES showing an emission maxima at ~ 2.1 eV. **c)** Reflectivity data of bare R3B.ClO$_4$ in a cavity for the same concentration and film thickness of R3B.ClO$_4$-SMILES reported in the main paper. The angle resolved reflection shows no strong coupling to the excitonic transition marked by the black dashed line resulting in weak coupling to the cavity mode.

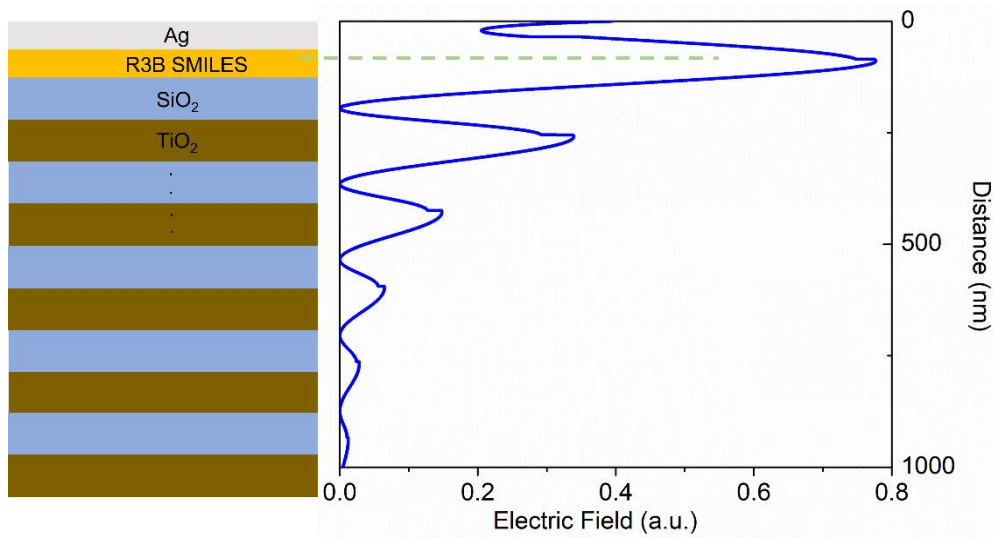

**Figure S2.** Electric field calculations of the completed Tamm cavity structure. The R3B SMILES film region overlaps with the maxima of the simulated electric field resulting in strong coupling.

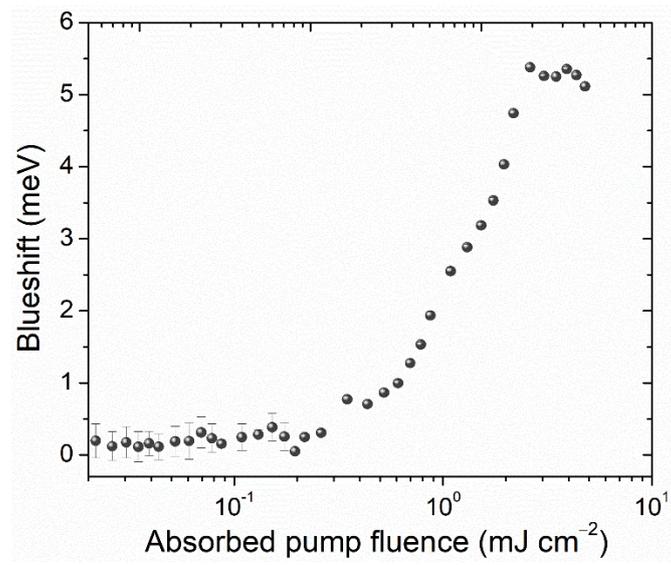

**Figure S3.** Blueshift of the lower polariton PL as a function of absorbed pump fluence.

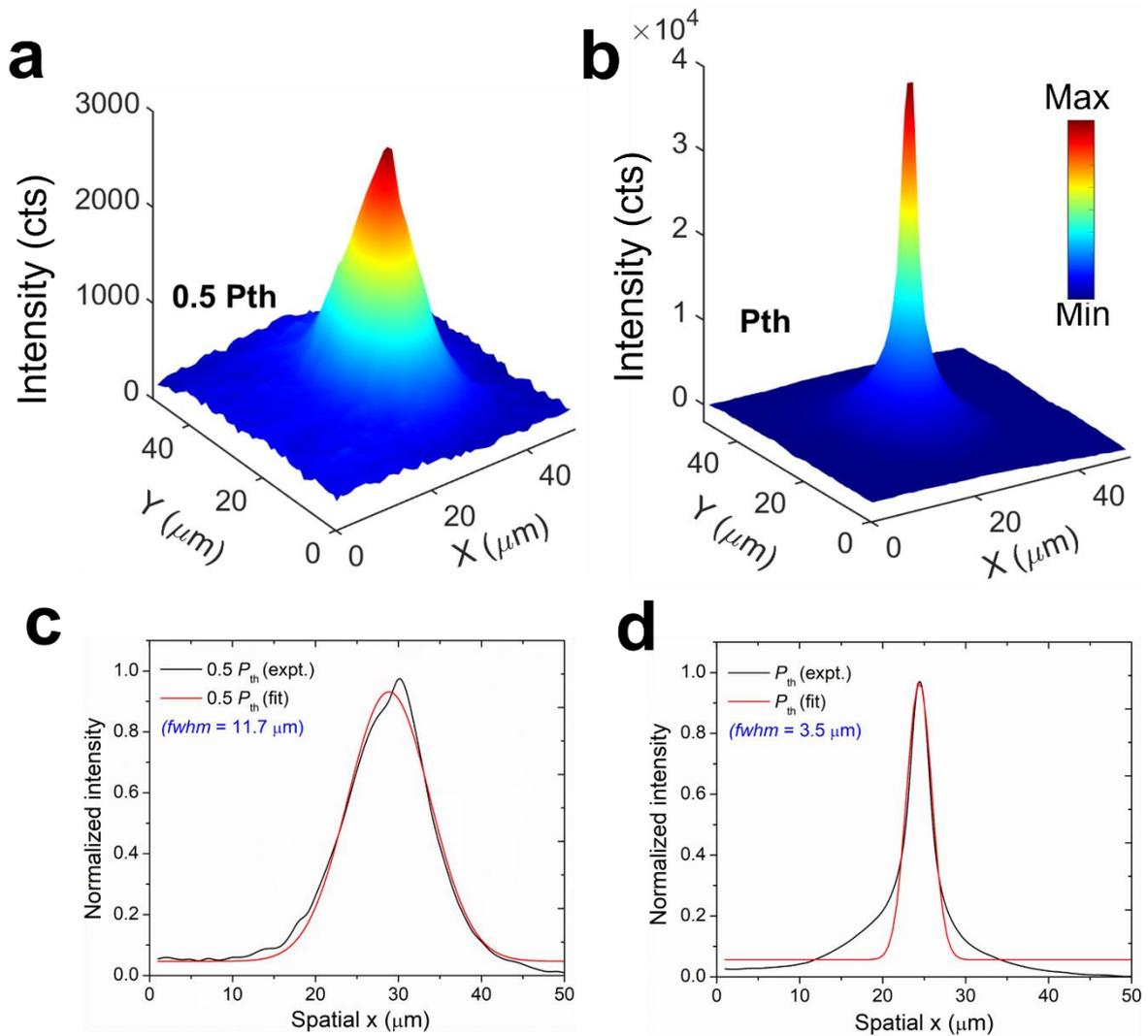

**Figure S4. a)** Below and **b)** At threshold laser fluence real space PL intensity surface maps showing enhancement in PL intensity and a narrowing of emission spot size. Corresponding linewidths for **c)** below and **d)** at threshold fluence are extracted by fitting a gaussian profile to the extracted linecuts of the emission spot size depicting narrowing of linewidth by more than factor of ~3 and a 16 fold emission intensity enhancement.

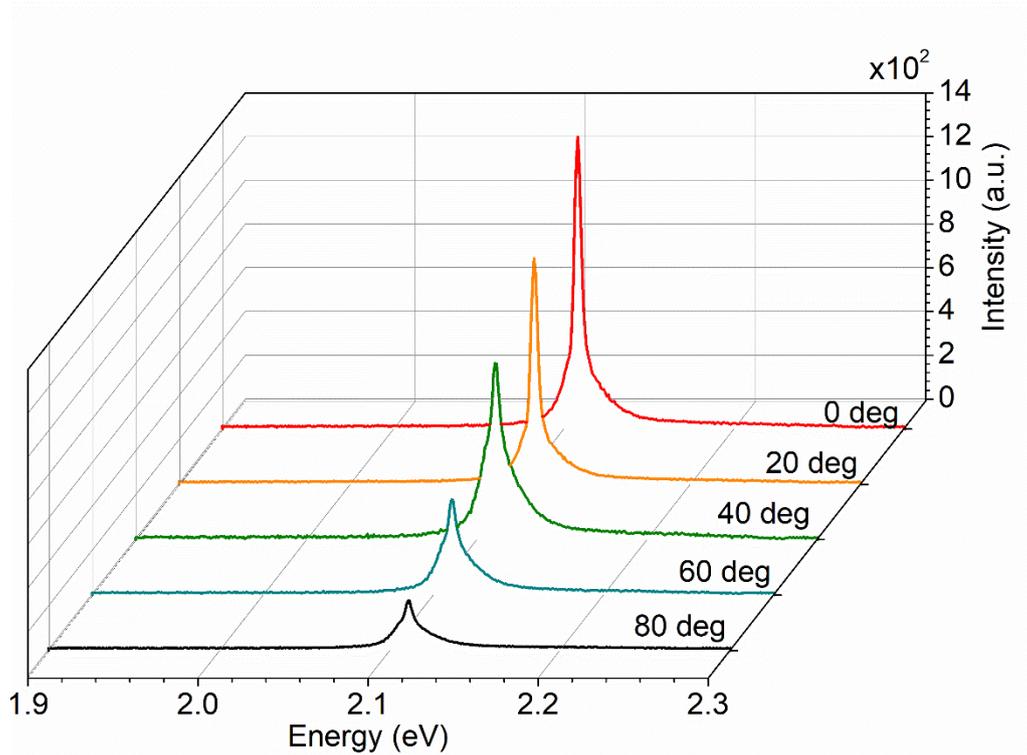

**Figure S5.** Spectra of polarization dependent microcavity PL of the condensate above threshold at 1.25 $P_{th}$ at different analyzer angles. The PL emission shows a strong polarization dependence with a 73% degree of linear polarization. As shown in the main text, the PL emission of the condensate below the threshold does not show any polarization dependence.

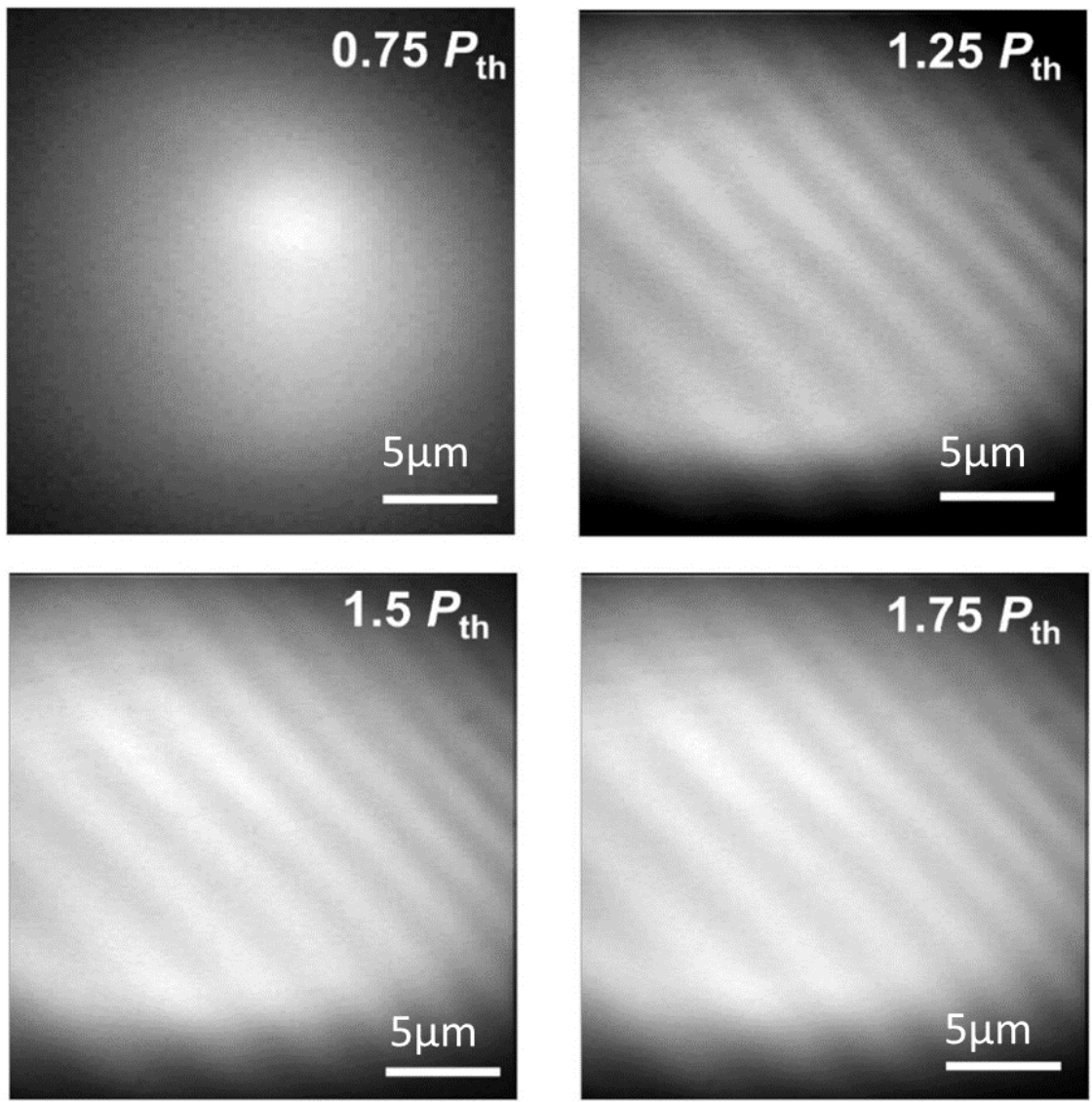

**Figure S6.** Real space PL interference maps in a Michelson interferometer configuration for detecting spatial coherence as a function of pump fluence. The occurance of fringes in the interferogram above the threshold power indicates the appearance of phase coherence in the strongly coupled condensate system.